\documentclass[epsf]{aa}
\usepackage{natbib}
\usepackage{graphicx}

\def\cite#1{\citealt{#1}}

\def\submitted{submitted}

\def\astroph#1{ (astro-ph/#1)}

\def\ExA{Experimental Astronomy}

\def\NewAR{New Astron. Rev.}
\def\PhR{Phys. Rev.}

\def\ibvs{Inf. Bull. Variable Stars}

\def\PublisherCambridge{Cambridge: Cambridge University Press}
\def\PublisherKluwer{Dordrecht: Kluwer Academic Publishers}

\begin{document}

\title{The Nature of V359 Centauri Revealed: \\
       New Long-Period SU~UMa-Type Dwarf Nova}
\subtitle{}
\authorrunning{T. Kato et al.}
\titlerunning{SU UMa-Type Dwarf Nova V359 Centauri}

\author{Taichi Kato\inst{1}
        \and Rod Stubbings\inst{2}
        \and Peter Nelson\inst{3}
        \and Roland Santallo\inst{4}
        \and Ryoko Ishioka\inst{1}
        \and Makoto Uemura\inst{1}
        \and Takahiro Sumi\inst{5,6}
        \and Yasushi Muraki\inst{5}
        \and Pam Kilmartin\inst{7}
        \and Ian Bond\inst{8}
        \and Sachiyo Noda\inst{5,9}
        \and Philip Yock\inst{10}
        \and John B. Hearnshaw\inst{7}
        \and Berto Monard\inst{11}
        \and Hitoshi Yamaoka\inst{12}
}

\institute{
  Department of Astronomy, Kyoto University, Kyoto 606-8502, Japan
  \and 19 Greenland Drive, Drouin 3818, Victoria, Australia
  \and RMB 2493, Ellinbank 3820, Australia
  \and Southern Stars Observatory, Po Box 60972, 98702 FAAA TAHITI,
       French Polynesia
  \and Solar-Terrestrial Environment Laboratory, Nagoya University,
       Nagoya 464-8601, Japan
  \and Department of Astrophysical Sciences, Princeton University,
       Peyton Hall, Princeton, NJ 08544, USA
  \and Department of Physics and Astronomy, University of Canterbury,
       Christchurch, New Zealand
  \and Inst. of Astronomy, University of Edinburgh, Royal Observatory,
       Edinburgh, UK
  \and Astronomical Data Analysis Center, National Astronomical Observatory
       of Japan, Mitaka, Tokyo 181-8588, Japan
  \and Department of Physics, University of Auckland, Auckland, New Zealand
  \and Bronberg Observatory, PO Box 11426, Tiegerpoort 0056, South Africa
  \and Faculty of Science, Kyushu University, Fukuoka 810-8560, Japan
}

\offprints{Taichi Kato, \\ e-mail: tkato@kusastro.kyoto-u.ac.jp}

\date{Received / accepted }

\abstract{
  We detected four outbursts of V359 Cen (possible nova discovered in 1939)
between 1999 and 2002.  Time-resolved CCD photometry during two outbursts
(1999 and 2002) revealed that V359 Cen is actually a long-period SU UMa-type
dwarf nova with a mean superhump period of 0.08092(1) d.  We identified
its supercycle length as 307--397 d.  This secure identification of the
superhump period precludes the previously supposed possibility that V359 Cen
could be related to a WZ Sge-type system with a long persistence of
late superhumps.  The outburst characteristics of V359 Cen are, however,
rather unusual in its low occurrence of normal outbursts.
}

\maketitle

\section{Introduction}

  Cataclysmic variables (CVs) are close binary systems consisting of
a white dwarf and a red dwarf secondary transferring matter via the Roche
lobe overflow (for a review of CVs, see \cite{war95book}).  CVs are
subdivided into several categories, including dwarf novae (DNe) and
novae.  Both DNe and novae are characterized by the presence of
a sudden increase of brightness (outburst).  Although the mechanisms
of DN-type outbursts (cf. \cite{osa96review}) and nova outbursts
(cf. \cite{sta87novareview,sta99novareview,sta00novareview})
are different, observational discrimination between rarely outbursting
DNe and novae can be sometimes difficult (see \cite{dow81wzsge} and
\cite{kat01hvvir} for classical and recent examples, respectively).
Since rarely outbursting DNe can be easily confused with very fast novae,
these confusions may have skewed our statistical view of classical novae
\citep{dow86novadensity,lil87novarate,sha97novarate}.

   A large fraction of such confusions turned out to be SU UMa-type dwarf
novae or WZ Sge-type dwarf novae \citep{kat01hvvir}.  SU UMa-type
dwarf novae are a subclass of DNe.  WZ Sge-type dwarf novae are still
enigmatic, both in theory and to observations, SU UMa-type dwarf novae,
which very infrequently (once in $\sim$10 yr) show large-amplitude
($\sim$8 mag) outbursts \citep{bai79wzsge,dow81wzsge,pat81wzsge,odo91wzsge}.
All SU UMa-type dwarf novae, including WZ Sge-type dwarf novae, show
superhumps during their long, bright outbursts (superoutbursts).
[For a recent review of dwarf novae and SU UMa-type dwarf novae, see
\citet{osa96review} and \citet{war95suuma}, respectively.]  Superhumps
have periods (superhump period: $P_{\rm SH}$) a few percent longer than
the orbital periods ($P_{\rm orb}$)
\citep{vog80suumastars,war85suuma}, which is believed to be
a consequence of the apsidal motion \citep{osa85SHexcess,mol92SHexcess}
of a tidally induced eccentric accretion disk
\citep{whi88tidal,hir90SHexcess,lub91SHa}.  WZ Sge-type dwarf novae
are known to show a different kind of (super)humps during the
earliest stage of superoutbursts
\citep{kat96alcom,mat98egcnc,ish02wzsgeletter,osa02wzsgehump,kat02wzsgeESH}.
These (super)humps in WZ Sge-type dwarf novae have periods close to
$P_{\rm orb}$, which can be easily distinguished from usual SU UMa-type
superhumps.
The presence of superhumps thus provides a powerful photometric tool in
discriminating novae and SU UMa-type/WZ Sge-type dwarf novae once
an object undergoes another outburst.

   V359 Cen was originally discovered as a possible nova by A. Opolski
(see \cite{due87novaatlas}).  The object was visible on 19 plates taken
between 1939 April 20 and 27, and the recorded maximum was m$_{pg}$ = 13.8
\citep{due87novaatlas}.  After an examination of Harvard plates
of the corresponding epoch and Opolski's finding chart,
\citet{due87novaatlas} suggested a 21.0 mag quiescent counterpart.
The true nature of the object, however, remained uncertain.  The object
was even proposed to be a nova in the Galactic halo.  From distant
nova candidates, \citet{kat01hvvir} selected V359 Cen as a candidate
for a rarely outbursting dwarf nova.  A finding chart of the proposed
quiescent counterpart was presented in \citet{due87novaatlas}.

   \citet{mun98CVspec5} tried to study the proposed quiescent counterpart
spectroscopically, but the attempt failed because of its faintness
($V$ fainter than 20.5).  \citet{gil98v359cen} obtained a deep image
around V359 Cen, and showed that the profile is indistinguishable from
that of a normal star; there was no evidence of a nova shell.

   The situation dramatically changed when one of the authors (Rod Stubbings)
detected the second historical outburst on 1999 July 13
(vsnet-alert 3216).\footnote{
http://www.kusastro.kyoto-u.ac.jp/vsnet/Mail/alert3000/\\msg00216.html.
}  The object further underwent outbursts in 2000 May, 2001 April and
2002 June.  We photometrically observed two outbursts (1999 July and
2002 June) and revealed that V359 Cen is an SU UMa-type dwarf nova.
\citet{wou01v359cenxzeriyytel} obtained time-resolved CCD photometry
following the 1999 July outburst and detected a periodicity of 0.0779 d
(112 min), but interpretation of this period remained rather uncertain.

\section{Observations}

   The 1999 observation by the MOA team was performed using a 61 cm
Ritchey-Chr\'{e}tien Cassegrain telescope (f/6.25) with the MOA-cam2
\citep{yan00MOAcam2}, constructed with three SITe back-illuminated CCDs
(2047$\times$4095 pixels).
The MOA blue filter (MOA B) covers 395--620 nm and MOA red filter
covers 620--1050 nm.  The exposure times were 300 and 180 s for the 1999
July 14 and 15 data, respectively.
The magnitudes of the object
were measured with Dophot package.  The absolute calibration of the
magnitudes was done using an ensemble of $\sim$40 neighboring stars,
whose zero-point was determined using about 100 LMC standard stars measureed
with the Hubble Space Telescope (HST).  The MOA magnitudes can be linked to
the standard $V$ and $R_{\rm c}$ systems using Eq. \ref{equ:moaconv},
where red and blue denote MOA red and MOA blue magnitudes
\citep{nod02MOALMC}.  Since the blue and red observations were not
completely simultaneous, we list the magnitudes on the MOA photometric
system in Table \ref{tab:moadata}.

\begin{eqnarray}
V         & = & blue - 0.16(blue-red) + const_1 \nonumber \\
R_{\rm c} & = & red  + 0.29(blue-red) + const_2 \label{equ:moaconv}
\end{eqnarray}

   The 2002 observations were undertaken by the VSNET Collaboration.\footnote{
http://www.kusastro.kyoto-u.ac.jp/vsnet/.}  The equipment and reduction
software are summarized in Table \ref{tab:equipment}.  The Kyoto
observations were analyzed using the Java$^{\rm TM}$-based PSF photometry
package developed by one of the authors (TK).  The other observers performed
aperture photometry.  The magnitudes were given relative to
GSC 7750.220, whose constancy during the observation was confirmed
by a comparison with USNO-A1.0 0450.13739601.  All systems are close to
$R_{\rm c}$.
The journal of the 2002 observations are summarized in Table \ref{tab:log}.

   Barycentric corrections to the observed times were applied before the
following analysis.

\begin{table}
\caption{MOA photometric data.} \label{tab:moadata}
\begin{center}
\begin{tabular}{cccc}
\hline
BJD$-$2400000 & Filter & MOA mag & Error \\
\hline
51373.945931 & red & -12.354 & 0.023 \\
51373.950698 & blue & -12.551 & 0.028 \\
51373.955478 & red & -12.354 & 0.018 \\
51373.960247 & blue & -12.486 & 0.025 \\
51373.965014 & red & -12.262 & 0.017 \\
51373.969783 & blue & -12.502 & 0.015 \\
51373.974551 & red & -12.192 & 0.020 \\
51373.979319 & blue & -12.431 & 0.016 \\
51373.984088 & red & -12.136 & 0.020 \\
51373.988867 & blue & -12.399 & 0.019 \\
51373.993635 & red & -12.174 & 0.014 \\
51373.998404 & blue & -12.475 & 0.010 \\
51374.003171 & red & -12.171 & 0.012 \\
51374.007940 & blue & -12.491 & 0.009 \\
51374.773159 & blue & -12.616 & 0.030 \\
51374.776677 & blue & -12.607 & 0.027 \\
51374.780195 & blue & -12.580 & 0.027 \\
51374.783725 & blue & -12.551 & 0.031 \\
51374.787244 & blue & -12.520 & 0.031 \\
51374.790762 & blue & -12.520 & 0.024 \\
51374.794291 & blue & -12.519 & 0.025 \\
51374.797810 & blue & -12.528 & 0.033 \\
51374.801351 & blue & -12.473 & 0.040 \\
51374.804881 & blue & -12.479 & 0.032 \\
51374.809256 & blue & -12.487 & 0.029 \\
51374.812774 & blue & -12.473 & 0.020 \\
\hline
\end{tabular}
\end{center}
\end{table}

\begin{table}
\caption{Equipment of the 2002 CCD photometry.} \label{tab:equipment}
\begin{center}
\begin{tabular}{cccc}
\hline
Observer   & Telescope &  CCD  & Software \\
\hline
Nelson     & 32-cm Newtonian & ST-8E & AIP4Win \\
Santallo   & 20-cm SCT & ST-7E & AIP4Win \\
Monard     & 30-cm SCT & ST-7E & AIP4Win \\
Kyoto      & 30-cm SCT & ST-7E & Java$^a$ \\
\hline
 \multicolumn{4}{l}{$^a$ See text.} \\
\end{tabular}
\end{center}
\end{table}

\begin{table}
\caption{Journal of the 2002 CCD photometry.} \label{tab:log}
\begin{center}
\begin{tabular}{crccrc}
\hline
\multicolumn{2}{c}{2002 Date}& Start--End$^a$ & Exp(s) & $N$
        & Obs$^b$ \\
\hline
May  & 29 & 52423.932--52424.061 & 30 & 270 & N \\
     & 31 & 52425.773--52425.867 & 24 & 164 & S \\
June &  1 & 52426.786--52426.921 & 15 & 360 & S \\
     &  1 & 52426.883--52426.996 & 60 & 111 & N \\
     &  2 & 52427.851--52428.026 & 30 & 357 & N \\
     &  2 & 52427.870--52427.925 & 60 &  55 & S \\
     &  3 & 52428.918--52428.998 & 45 & 129 & N \\
     &  3 & 52428.954--52428.993 & 10 & 146 & K \\
     &  5 & 52430.743--52430.847 & 60 &  76 & S \\
     &  5 & 52430.954--52430.962 & 10 &  30 & K \\
     &  6 & 52431.954--52431.968 & 10 &  44 & K \\
     &  6 & 52432.182--52432.433 & 45 & 356 & M \\
\hline
 \multicolumn{6}{l}{$^a$ BJD$-$2400000.} \\
 \multicolumn{6}{l}{$^b$ N (Nelson), S (Santallo), K (Kyoto team), M (Monard)} \\
\end{tabular}
\end{center}
\end{table}

\begin{figure*}
  \begin{center}
  \includegraphics[angle=0,width=15cm]{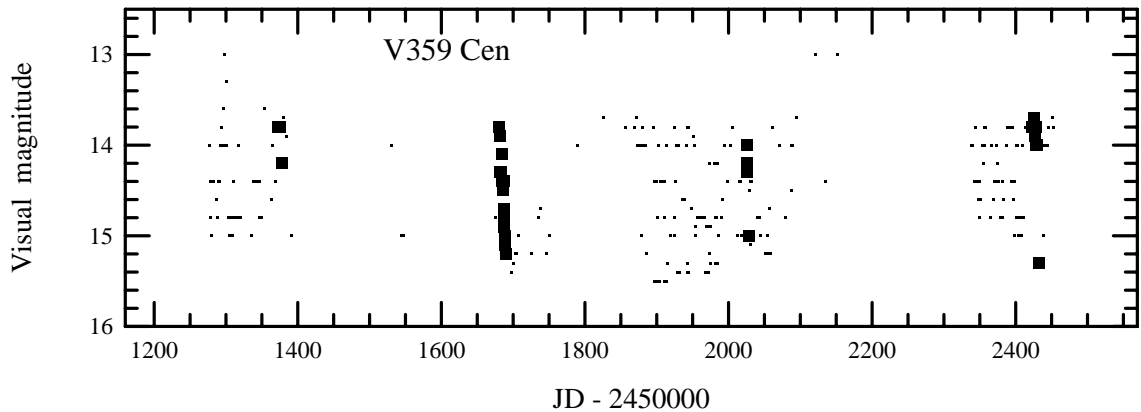}
  \end{center}
  \caption{Long-term visual light curve of V359 Cen constructed from the
  observations reported to the VSNET Collaboration.
  Large and small dot represent positive and
  negative (upper limit) observations, respectively.  Four outbursts
  are clearly seen.}
  \label{fig:long}
\end{figure*}

\section{Astrometry and Quiescent Counterpart}

   Astrometry of the outbursting V359 Cen was performed on CCD
images taken by R. Santallo (2002 June 1).  An average of measurements
of five images (GSC$-$2.2 system, about 20 reference stars; internal
dispersion of the measurements was $\sim$0$''$.05) has yielded
a position of 11$^h$ 58$^m$ 15$^s$.330, $-$41$^{\circ}$ 46$'$ 08$''$.44
(J2000.0).  The position agrees with the GSC$-$2.2 star at
11$^h$ 58$^m$ 15$^s$.322, $-$41$^{\circ}$ 46$'$ 08$''$.35
(epoch 1995.392 and magnitudes $r$ = 18.46, $b$ = 19.15)
and the USNO$-$A2.0 star at
11$^h$ 58$^m$ 15$^s$.330, $-$41$^{\circ}$ 46$'$ 09$''$.16
(epoch 1982.262 and magnitudes $r$ = 17.7, $b$ = 18.7).

   This identification confirms the quiescent magnitude
($V$ = 18.7) reported by \citet{wou01v359cenxzeriyytel}.
The quiescent magnitudes (21 or $V$ fainter than 20.5)
reported by \citet{due87novaatlas} and \citet{mun98CVspec5}
seem to be underestimated.  An examination of POSS I red plate
(limiting magnitude $R\sim$ 18.5)
shows that the object was near the detection limit
(presumably a result of a large air-mass).  This impression may
have affected the estimate by \citet{due87novaatlas}.

   The failure by \citet{mun98CVspec5} in obtaining a quiescent spectrum
is, however, difficult to reconcile with the value of $V$ = 18.7.
Since a few dwarf novae categorized to established or suspected SU UMa-type
dwarf novae are known to show high and low states in quiescence
(HT Cas: \cite{zha86htcas,woo95htcasXray,rob96htcas};
IR Com: \cite{ric95ircom,kat02ircom} and less established
BZ UMa: \cite{kal86bzumaiauc}).  Although it is still premature to draw
a firm conclusion, V359 Cen may belong to a small class of SU UMa-type
dwarf novae with high/low transitions in quiescence.

\section{Long-Term Light Curve}

   Fig. \ref{fig:long} shows the long-term visual light curve of V359 Cen
constructed from the observations reported to the VSNET Collaboration.
Large and small dot represent positive and negative (upper limit)
observations, respectively.  Four outbursts (1999 July, starting on
JD 2451373; 2000 May, on JD 2451680; 2001 April, on JD 2452025 and
2002 June, on JD 2452422) are clearly seen.  The intervals between
the detected outbursts are in the range of 307--397 d.  Although there
were unavoidable seasonal gaps in observations, these values seem to be
a representative outburst cycle length.  The observed maximum magnitudes
of the outbursts were $\sim$13.8.  This constancy of the maximum
magnitudes likely precludes the previously supposed possibility that
the maximum of the 1939 outburst was missed \citep{due87novaatlas}.

\section{The 1999 Outburst}

\begin{figure}
  \includegraphics[angle=0,width=8.8cm]{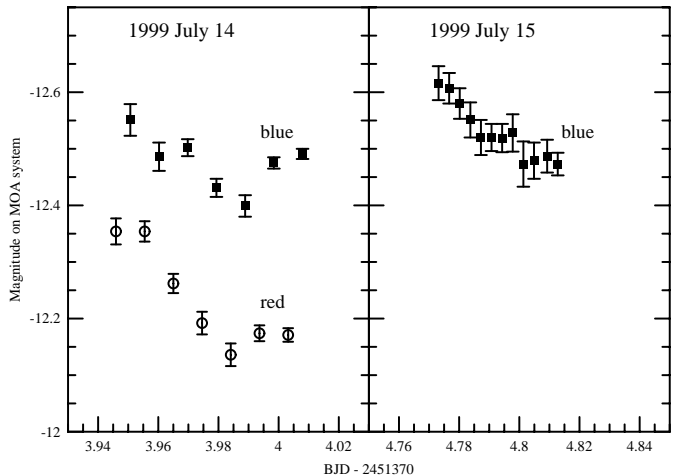}
  \caption{Superhumps of V359 Cen during the 1999 outburst.  The data
  are from the MOA observations.  Filled squares and open circles
  represent blue and red observation, respectively.
  The magnitudes are based on the MOA photometric system.}
  \label{fig:moalc}
\end{figure}

   Fig. \ref{fig:moalc} shows the enlarged light curve of V359 Cen
during the 1999 outburst.  The data are from the MOA observations.
Although complete phase coverage was impossible because of the
short available runs, the object clearly exhibited variations with
amplitudes of $\sim$0.15 mag, which can be attributed to superhumps.

\begin{figure}
  \includegraphics[angle=0,width=8cm]{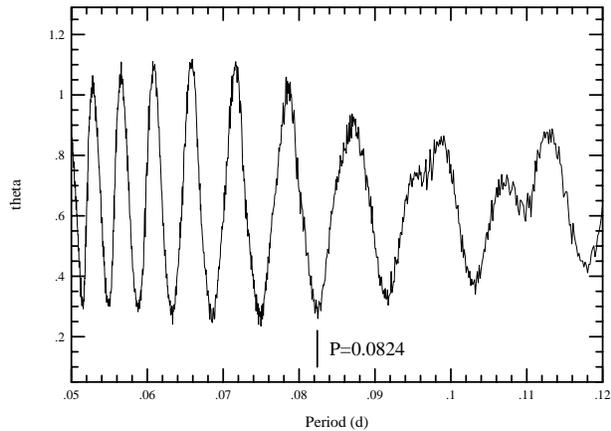}
  \caption{Period analysis of the MOA blue observations.  The denoted
  superhump period is the bast-selected alias based on the 2002
  observation.}
  \label{fig:moapdm}
\end{figure}

   A period analysis of the MOA blue data using Phase Dispersion
Minimization (PDM; \cite{PDM}) has yielded the theta diagram presented
in Fig. \ref{fig:moapdm}.  Although a unique alias selection is
impossible from these data only, we can safely choose the correct
alias of $P_{\rm SH}$ = 0.0824(4) d based on the later determination of
the superhump period (Sect. \ref{sec:Psh}).

\section{The 2002 Outburst}

\subsection{Course of the Outburst}

\begin{figure}
  \includegraphics[angle=0,width=8cm]{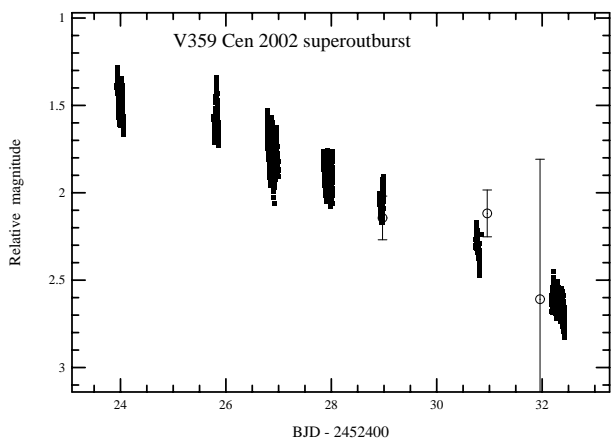}
  \caption{Light curve of the 2002 superoutburst of V359 Cen.  The magnitudes
  are relative to GSC 7750.220 and are on a system close to $R_{\rm c}$.
  Open circles with errors represent nightly averaged Kyoto observations.}
  \label{fig:overall}
\end{figure}

   The 2002 outburst was detected by Rod Stubbings on May 28 at
$m_{\rm vis}$ = 13.8 (vsnet-alert 7356).\footnote{
http://www.kusastro.kyoto-u.ac.jp/vsnet/Mail/alert7000/\\msg00356.html.
}  CCD time-resolved photometry started within a
day following this detection.  Fig. \ref{fig:overall} shows the
overall light curve of the 2002 superoutburst.  The magnitudes
are relative to GSC 7750.220 and are on a system close to $R_{\rm c}$.
After two days of the outburst, the system started to fade linearly
at a rate of 0.16 mag d$^{-1}$.  This slowly and linearly fading
phase (often referred to as {\it superoutburst plateau}) is very
characteristic of an SU UMa-type superoutburst
\citep{vog80suumastars,war85suuma}.

   The mean decline rate of 0.16 mag d$^{-1}$ during the plateau phase
is larger than those of other SU UMa-type dwarf novae
(Table \ref{tab:declinerate}) with
similar $P_{\rm SH}$ to that of V359 Cen (cf. Sect. \ref{sec:Psh}).
Since a higher mass-transfer rate from the secondary star tends to thermally
stabilizes the accretion disk and reduce the decay rate (e.g.
\cite{osa95eruma}), a rather exceptionally large decay rate in V359 Cen
may be a result of a systematically smaller mass-transfer rate.

\begin{table}
\caption{Mean Decline Rates of SU UMa-Type Dwarf Novae.}\label{tab:declinerate}
\begin{center}
\begin{tabular}{cccc}
\hline\hline
Object & $P_{\rm SH}$ (d) & Mean rate$^a$ & Ref. \\
\hline
HV Aur & 0.0855 & 0.035 & 1 \\
TU Crt & 0.0854 & 0.092 & 2 \\
AW Gem & 0.0794 & 0.08  & 3 \\
TT Boo & 0.0781 & 0.11  & 4 \\
\hline
\end{tabular}
\end{center}
{\footnotesize
$^a$ Mean decline rate (mag d$^{-1}$) during the superoutburst plateau. \\
{\bf References:}
  1: \citet{nog95hvaur},
  2: \citet{men98tucrt},
  3: \citet{kat96awgem},
  4: \citet{kat95ttboo}.
}
\end{table}

\subsection{Superhump Period and Evolution}\label{sec:Psh}

\begin{figure}
  \includegraphics[angle=0,width=8cm,height=11cm]{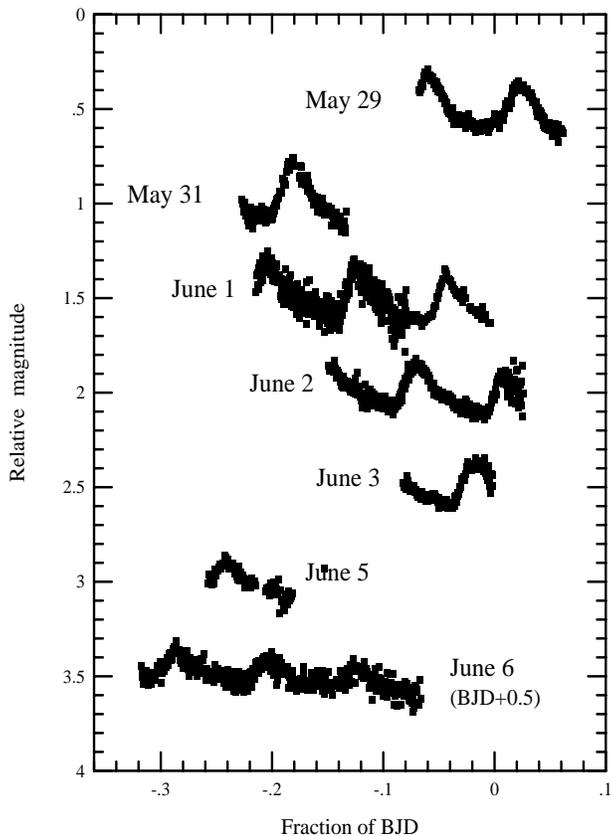}
  \caption{Nightly light curves.  Superhumps are clearly visible on
  all nights.}
  \label{fig:night}
\end{figure}

\begin{figure}
  \includegraphics[angle=0,width=8cm]{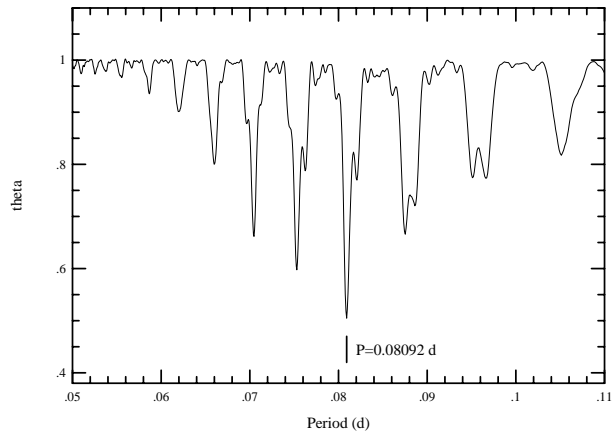}
  \caption{Period analysis of the 2002 data between May 31 and June 6
  (superoutburst plateau).
  The linear declining trend has been subtracted before the analysis.
  }
  \label{fig:pdm}
\end{figure}

\begin{figure}
  \includegraphics[angle=0,width=8cm]{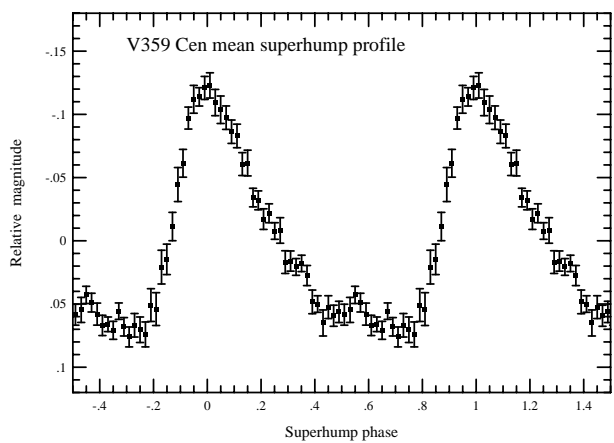}
  \caption{Mean superhump profile of V359 Cen.}
  \label{fig:sh}
\end{figure}

   Fig. \ref{fig:night} shows enlarged nightly light curves.
Superhumps are clearly visible on all observed nights.
Fig. \ref{fig:pdm} shows the result of a PDM period analysis of the
2002 data between May 31 and June 6 (superoutburst plateau).
The linear declining trend has been subtracted before the analysis.
The resultant best $P_{\rm SH}$ is 0.08092(1) d.  The selection of the
correct alias has been confirmed by independent period analyses of
individual long continuous runs.  This period established that V359 Cen
is a long-period SU UMa-type dwarf nova.  Fig. \ref{fig:sh} shows the
phase-averaged profile of superhumps.  The rapidly rising and slowly
fading superhump profile is characteristic to an SU UMa-type
dwarf nova \citep{vog80suumastars,war85suuma}.

   We extracted the maxima times of superhumps from the light curve by eye.
The averaged times of a few to several points close to the maximum were
used as representatives of the maxima times.  The errors of the maxima
times are usually less than $\sim$0.002 d.  The resultant superhump maxima
are given in Table \ref{tab:shmax}.  The values are given to 0.0001 d in
order to avoid the loss of significant digits in a later analysis.
The cycle count ($E$) is defined as the cycle number since BJD 2452423.939.
A linear regression to the observed superhump times gives the following
ephemeris:

\begin{equation}
{\rm BJD (maximum)} = 2452423.9503 + 0.08108 E. \label{equ:reg1}
\end{equation}

\begin{figure}
  \includegraphics[angle=0,width=8cm]{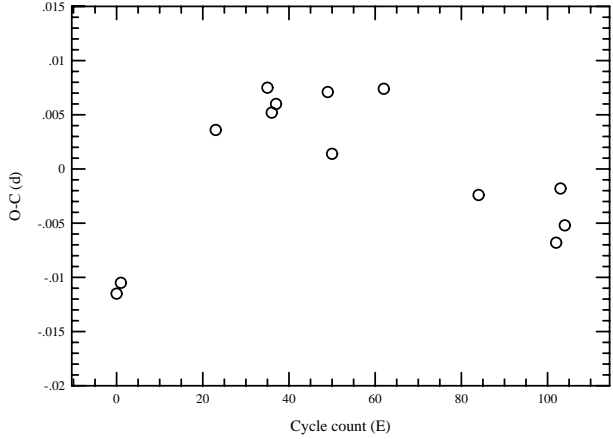}
  \caption{$O-C$ diagram of superhump maxima.  The $O-C$'s are calculated
  against equation \ref{equ:reg1}.}
  \label{fig:oc}
\end{figure}

   Fig. \ref{fig:oc} shows the ($O-C$)'s against the mean superhump period
(0.08108 d) from Eq. \ref{equ:reg1}.  Although the general trend
can be expressed by a negative quadratic term of
$\dot{P}$ = $-10.8(1.5) \times 10^{-6}$ d cycle$^{-1}$, or
$\dot{P}/P$ = $-13.3(1.9) \times 10^{-5}$, superhump maxima with $E \geq 23$
can be well expressed by a constant period of $P$ = 0.08094 d and
$|O-C|$'s less than 0.005 d.  This finding indicates that the superhump
period was virtually constant during the plateau phase (the nominal
$\dot{P}$ during this period os $-5.3(2.8) \times 10^{-6}$ d cycle$^{-1}$).
A sudden change between $E$ = 0 and $E$ = 23 can be interpreted as a result
of rapid evolution of superhumps during the earliest stage of the
superoutburst.  The ``textbook" evolutionary time-scales (2--3 d) of
superhumps in long-period SU UMa-type dwarf novae \citep{war85suuma} also
support this interpretation.  The large period change of $P_{\rm SH}$
observed during the earliest stage of the 2002 superoutburst may explain
the slight discrepancy of the periods between the 1999 and 2002
observations (the 1999 observation corresponds to the earliest stage
of a superoutburst).  The period determined from $0 \leq E \leq 23$
observations is 0.0817 d, sufficiently longer than the most likely orbital
period (Sect. \ref{sec:Porb}), precludes the possibility of WZ Sge-type
early (super)humps as the origin of these early modulations.

\begin{table}
\caption{Times of superhump maxima.}\label{tab:shmax}
\begin{center}
\begin{tabular}{ccc}
\hline\hline
$E^a$  & BJD$-$2400000 & $O-C^b$ \\
\hline
   0 & 52423.9388 & -0.0115 \\
   1 & 52424.0209 & -0.0105 \\
  23 & 52425.8188 &  0.0036 \\
  35 & 52426.7956 &  0.0075 \\
  36 & 52426.8744 &  0.0052 \\
  37 & 52426.9563 &  0.0060 \\
  49 & 52427.9303 &  0.0071 \\
  50 & 52428.0057 &  0.0014 \\
  62 & 52428.9847 &  0.0074 \\
  84 & 52430.7587 & -0.0024 \\
 102 & 52432.2138 & -0.0068 \\
 103 & 52432.2999 & -0.0018 \\
 104 & 52432.3774 & -0.0052 \\
\hline
 \multicolumn{3}{l}{$^a$ Cycle count since BJD 2452423.939.} \\
 \multicolumn{3}{l}{$^b$ $O-C$ calculated against Eq.
                    \ref{equ:reg1}.} \\
\end{tabular}
\end{center}
\end{table}

\subsection{Super-QPOs}

\begin{figure}
  \includegraphics[angle=0,width=8.8cm]{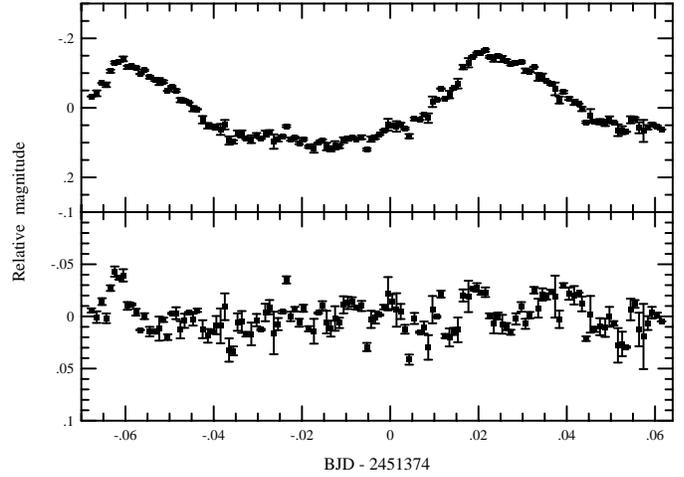}
  \caption{(Upper) Enlarged light curve on May 29.  (Lower) Residuals
  of the May 29 light curve after subtracting the superhump signal
  and a slow linear trend.  Small-amplitude modulations with a typical
  time scale of $\sim$0.02 d are present.
  }
  \label{fig:qpolc}
\end{figure}

\begin{figure}
  \includegraphics[angle=0,width=8cm]{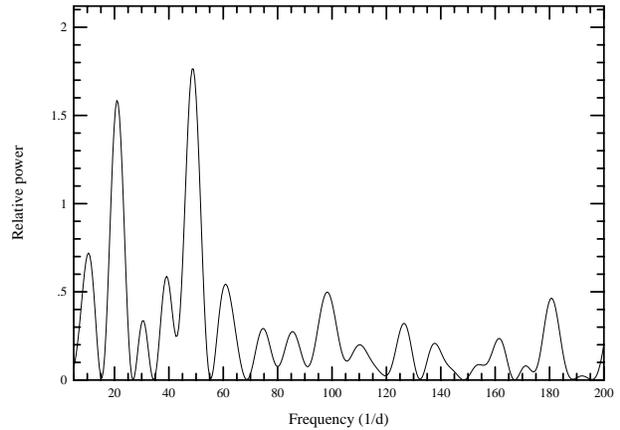}
  \caption{Power spectrum of the QPO signal.  The strongest power is
  present at a frequency of $\sim$49 d$^{-1}$, which corresponds to
  a period of $\sim$0.02 d.
  }
  \label{fig:qpopow}
\end{figure}

   Some SU UMa-type dwarf novae are known to show large-amplitude,
highly coherent quasi-periodic oscillations (QPOs) during the evolution
stage of superhumps \citep{kat92swumasuperQPO,kat02efpeg}.
These QPOs are sometimes referred to as ``super-QPOs".
The light curve of V359 Cen on May 29 shows a hint of such QPOs.
The lower panel of Fig. \ref{fig:qpolc} shows residuals
of the May 29 light curve after subtracting the superhump signal
by using a Fourier decomposition of the superhump profile up to the third
harmonics, and subtracting a slow linear trend.  Small-amplitude modulations
with a typical time scale of $\sim$0.02 d are present.
Fig. \ref{fig:qpopow} shows a power spectrum of the residual shown
in the lower panel of Fig. \ref{fig:qpolc}.  The strongest power is
present at a frequency of $\sim$49 d$^{-1}$, which corresponds to
a period of $\sim$0.02 d.

   Although the QPOs were not as prominent as seen in SW UMa
\citep{kat92swumasuperQPO} or \citep{kat02efpeg}, the absence of similar
signals on later nights suggest that this variation is a kind of
super-QPOs.  As shown in Sect. \ref{sec:Psh}, the epoch of the detection
of QPOs corresponds to the rapidly evolving stage of superhumps.
This finding further supports an idea that super-QPOs are associated
with the growth of superhumps \citep{kat92swumasuperQPO,kat02efpeg}

\section{V359 Cen as an SU UMa-Type Dwarf Nova}

\subsection{Outburst Characteristics}

   Table \ref{tab:outburst} lists the observed parameters of the four
recorded outbursts.  All the recorded outbursts have durations longer
than 4 d, which indicate that all the recorded outbursts were superoutbursts.
The intervals between these outbursts (307--397 d) are thus regarded
as the supercycle of this SU UMa-type dwarf nova.
Since the present duty cycle of observations is about 50\%, there
remains a small possibility that the true supercycle is the half of this
value.  Further dense monitoring is strongly encouraged to completely
exclude this possibility.
It is well established
that almost all well-studied SU UMa-type dwarf novae with similar supercycle
lengths show a few to $\sim$ten normal outburst within one supercycle
\citep{war95suuma,nog97sxlmi}.  Since normal outbursts of SU UMa-type
dwarf novae are usually only 0.5--1.0 mag fainter than superoutbursts
\citep{war85suuma}, some of such normal outbursts should have been
detected.  The present non-detection of normal outbursts seems to
suggest that V359 Cen has fewer normal outbursts than in usual SU UMa-type
dwarf novae.  Although future deeper monitoring for normal outbursts is
absolutely needed, the occurrence of normal outbursts in V359 Cen may be
effectively suppressed by an unknown mechanism; similar instances have
been reported in other SU UMa-type dwarf novae
\citep{kat01sxlmi,kat02v344lyr}.  If normal outbursts are entirely missing,
V359 Cen would become the first long-period analog of V844 Her
which has been known to show only superoutbursts (\cite{kat00v844her}:
$P_{\rm SH}$ = 0.05592 d, supercycle lengths = 220--290 d).
Similar SU UMa-type dwarf novae with predominance of superoutbursts
are highly concentrated in the short period region
(cf. \cite{kat00v844her}).  V359 Cen may be an exceptional object in
its combination of outburst characteristics and the superhump (or orbital)
period.

   We also note that the outburst characteristics of V359 Cen also
resemble those of long-period SU UMa-type dwarf novae EF Peg
\citep{how93efpeg,kat02efpeg} and V725 Aql \citep{uem01v725aql}.
Both EF Peg and V725 Aql only infrequently show normal outbursts,
which is exceptional among long-$P_{\rm orb}$ SU UMa-type dwarf novae
(cf. \cite{war95suuma}).  While long outburst recurrence times imply
that these systems have low mass-transfer rates
\citep{ich94cycle,osa96review}, recent detailed calculations of
the evolution of CVs (e.g. \cite{pod01amcvn}) suggest that mass-transfer
rates are higher in long-period systems even if the effect of stellar
core evolution is properly taken into account.  These systems (EF Peg,
V725 Aql and possibly V359 Cen) may be violating the modern evolutionary
scenario of CVs.  Future determination of the binary parameters and
stellar composition analysis of these systems are therefore strongly
encouraged.

\subsection{Superhump Excess and Late Superhumps}\label{sec:Porb}

   \citet{wou01v359cenxzeriyytel} tentatively identified their photometric
period (112 min) as late superhumps
\citep{hae79lateSH,vog83lateSH,vanderwoe88lateSH,hes92lateSH},
which are known to have similar periods with ordinary superhumps
(i.e. a few percent longer than $P_{\rm orb}$), but have phases of
$\sim$ 0.5 different from those of ordinary superhumps.  Since this
signal was observed long after then cessation of the 1999 outburst,
\citet{wou01v359cenxzeriyytel} suggested that V359 Cen may have shown
a long persistence of late superhumps as was observed in a WZ Sge-type
dwarf nova in EG Cnc \citep{kat97egcnc,pat98egcnc}.

   The present correct identification of the superhump period, however,
indicates that the 112 min periodicity observed by
\citet{wou01v359cenxzeriyytel} can not be attributed to superhumps,
but can be better understood to represent $P_{\rm orb}$.  This observation,
on the contrary to the suggestion by \citet{wou01v359cenxzeriyytel},
indicates that the superhumps or late superhumps in V359 Cen must
have decayed more rapidly, as in other usual SU UMa-type dwarf novae
(see \citet{kat01wxcet} for a recent example of the decay of late
superhumps).  V359 Cen is thus unlikely related to WZ Sge-type dwarf
novae which always show long persistence of late superhumps.

   By adopting $P_{\rm orb}$ = 0.0779 d, we obtain a fractional superhump
excess ($\epsilon=P_{\rm SH}/P_{\rm orb}-1$) of 3.9\% for the
best $P_{\rm SH}$ is 0.08092(1) d (cf. Sect. \ref{sec:Psh}).
Using the ``mean" $P_{\rm SH}$ = 0.08108 d from the entire 2002
superoutburst, we obtain $\epsilon$ = 4.1\%.  These fractional superhump
excesses are not unusual for an SU UMa-type dwarf nova with this
$P_{\rm orb}$ \citep{mol92SHexcess,pat98evolution}, suggesting that
V359 Cen should have a normal binary mass ratio $q=M_2/M_1$ in spite of
its rather unusual outburst characteristics.

\begin{table}
\caption{List of Outbursts.} \label{tab:outburst}
\begin{center}
\begin{tabular}{cccc}
\hline
JD start$^a$ & JD end$^a$ & Maximum & Duration (d) \\
\hline
51372.9 & 51377.9 & 13.8 & $>$5 \\
51680.0 & 51690.0 & 13.8 & $>$10 \\
52025.0 & 52029.0 & 14.0 & $>$4 \\
52422.9 & 52433.0 & 13.8 & $>$10 \\
\hline
 \multicolumn{4}{l}{$^a$ JD$-$2400000.} \\
\end{tabular}
\end{center}
\end{table}

\section{Summary}

   We detected four outbursts of V359 Cen (possible nova discovered in 1939)
between 1999 and 2002.  Time-resolved CCD photometry during two outbursts
(1999 and 2002) revealed that V359 Cen is actually a long-period SU UMa-type
dwarf nova with a mean superhump period of 0.08092(1) d.  We identified
its supercycle length as 307--397 d.  This secure identification of the
superhump period precludes the previously supposed possibility that V359 Cen
could be related to a WZ Sge-type system with a long persistence of
late superhumps.  The outburst characteristics of V359 Cen are, however,
rather unusual in its low occurrence of normal outbursts.  The fractional
superhump excess is 3.9--4.1\%, which suggests that V359 Cen should have
a normal binary mass ratio in spite of its rather unusual outburst
characteristics.  We also obtained a secure identification of the quiescent
counterpart and discussed on the possibility of high/low state changes.
The evolution of superhumps and their period change was closely followed.
We also detected super-QPO-type variation (period $\sim$0.02 d) during
the earliest stage of the 2002 superoutburst.

\vskip 3mm

The authors are grateful to Andrew Pearce who reported
visual observations of V359 Cen to VSNET.
This work is partly supported by a grant-in aid (13640239) from the
Japanese Ministry of Education, Culture, Sports, Science and Technology.
Part of this work is supported by a Research Fellowship of the
Japan Society for the Promotion of Science for Young Scientists (MU).
The CCD operation of the Bronberg Observatory is partly sponsored by
the Center for Backyard Astrophysics.
The CCD operation by Peter Nelson is on loan from the AAVSO,
funded by the Curry Foundation.
This research has made use of the Digitized Sky Survey producted by STScI, 
the ESO Skycat tool, the VizieR catalogue access tool, and
the USNOFS Image and Catalogue Archive operated by the United States Naval
Observatory, Flagstaff Station (http://www.nofs.navy.mil/data/fchpix/).


\begin{thebibliography}{62}
\expandafter\ifx\csname natexlab\endcsname\relax\def\natexlab#1{#1}\fi

\bibitem[{Bailey(1979)}]{bai79wzsge}
Bailey, J. 1979, \mnras, 189, 41P

\bibitem[{Downes(1986)}]{dow86novadensity}
Downes, R.~A. 1986, \apj, 307, 170

\bibitem[{Downes \& Margon(1981)}]{dow81wzsge}
Downes, R.~A. \& Margon, B. 1981, \mnras, 197, 35P

\bibitem[{Duerbeck(1987)}]{due87novaatlas}
Duerbeck, H.~W. 1987, \ssr, 45, 1

\bibitem[{Gill \& O'Brien(1998)}]{gil98v359cen}
Gill, C.~D. \& O'Brien, T.~J. 1998, \mnras, 300, 221

\bibitem[{Haefner {et~al.}(1979)Haefner, Schoembs, \& Vogt}]{hae79lateSH}
Haefner, R., Schoembs, R., \& Vogt, N. 1979, \aap, 77, 7

\bibitem[{Hessman {et~al.}(1992)Hessman, Mantel, Barwig, \&
  Schoembs}]{hes92lateSH}
Hessman, F.~V., Mantel, K.-H., Barwig, H., \& Schoembs, R. 1992, \aap, 263, 147

\bibitem[{Hirose \& Osaki(1990)}]{hir90SHexcess}
Hirose, M. \& Osaki, Y. 1990, \pasj, 42, 135

\bibitem[{Howell {et~al.}(1993)Howell, Schmidt, DeYoung, Fried, Schmeer, \&
  Gritz}]{how93efpeg}
Howell, S.~B., Schmidt, R., DeYoung, J.~A., {et~al.} 1993, \pasp, 105, 579

\bibitem[{Ichikawa \& Osaki(1994)}]{ich94cycle}
Ichikawa, S. \& Osaki, Y. 1994, in Theory of Accretion Disks-2, ed. W.~J.
  Duschl, J.~Frank, F.~Meyer, E.~Meyer-Hofmeister, \& W.~M. Tscharnuter
  (\PublisherKluwer), 169

\bibitem[{Ishioka {et~al.}(2002)Ishioka, Uemura, Matsumoto, Ohashi, Kato, Masi,
  Novak, Pietz, Martin, Starkey, Kiyota, Oksanen, Moilanen, Cook, Kral, Hynek,
  Kolasa, Vanmunster, Richmond, Kern, Davis, Crabtree, Beaulieu, Davis,
  Aggleton, Gazeas, Niarchos, Yushchenko, Mallia, Fiaschi, Good, Boyd, Sano,
  Morikawa, Moriyama, Mennickent, Arenas, Ohshima, \&
  Watanabe}]{ish02wzsgeletter}
Ishioka, R., Uemura, M., Matsumoto, K., {et~al.} 2002, \aap, 381, L41

\bibitem[{Kaluzny(1986)}]{kal86bzumaiauc}
Kaluzny, J. 1986, \iaucirc, 4287

\bibitem[{Kato(1995)}]{kat95ttboo}
Kato, T. 1995, \ibvs, 4243

\bibitem[{Kato(1996)}]{kat96awgem}
---. 1996, \pasj, 48, 777

\bibitem[{Kato(2001)}]{kat01sxlmi}
---. 2001, \ibvs, 5071

\bibitem[{Kato(2002{\natexlab{a}})}]{kat02wzsgeESH}
---. 2002{\natexlab{a}}, \pasj, 54, L11

\bibitem[{Kato(2002{\natexlab{b}})}]{kat02efpeg}
---. 2002{\natexlab{b}}, \pasj, 54, 87

\bibitem[{Kato {et~al.}(2002{\natexlab{a}})Kato, Baba, \& Nogami}]{kat02ircom}
Kato, T., Baba, H., \& Nogami, D. 2002{\natexlab{a}}, \pasj, 54, 79

\bibitem[{Kato {et~al.}(1992)Kato, Hirata, \& Mineshige}]{kat92swumasuperQPO}
Kato, T., Hirata, R., \& Mineshige, S. 1992, \pasj, 44, L215

\bibitem[{Kato {et~al.}(2001{\natexlab{a}})Kato, Matsumoto, Nogami, Morikawa,
  \& Kiyota}]{kat01wxcet}
Kato, T., Matsumoto, K., Nogami, D., Morikawa, K., \& Kiyota, S.
  2001{\natexlab{a}}, \pasj, 53, 893

\bibitem[{Kato {et~al.}(1996)Kato, Nogami, Baba, Matsumoto, Arimoto, Tanabe, \&
  Ishikawa}]{kat96alcom}
Kato, T., Nogami, D., Baba, H., {et~al.} 1996, \pasj, 48, L21

\bibitem[{Kato {et~al.}(1997)Kato, Nogami, Matsumoto, \& Baba}]{kat97egcnc}
Kato, T., Nogami, D., Matsumoto, K., \& Baba, H. 1997, preprint, 
  ftp://vsnet.kusastro.kyoto-u.ac.jp/pub/\\vsnet/preprints/EG\_Cnc/

\bibitem[{Kato {et~al.}(2002{\natexlab{b}})Kato, Poyner, \&
  Kinnunen}]{kat02v344lyr}
Kato, T., Poyner, G., \& Kinnunen, T. 2002{\natexlab{b}}, \mnras, 330, 53

\bibitem[{Kato {et~al.}(2001{\natexlab{b}})Kato, Sekine, \&
  Hirata}]{kat01hvvir}
Kato, T., Sekine, Y., \& Hirata, R. 2001{\natexlab{b}}, \pasj, 53, 1191

\bibitem[{Kato \& Uemura(2000)}]{kat00v844her}
Kato, T. \& Uemura, M. 2000, \ibvs, 4902

\bibitem[{Liller \& Mayer(1987)}]{lil87novarate}
Liller, W. \& Mayer, B. 1987, \pasp, 99, 606

\bibitem[{Lubow(1991)}]{lub91SHa}
Lubow, S.~H. 1991, \apj, 381, 259

\bibitem[{Matsumoto {et~al.}(1998)Matsumoto, Nogami, Kato, \&
  Baba}]{mat98egcnc}
Matsumoto, K., Nogami, D., Kato, T., \& Baba, H. 1998, \pasj, 50, 405

\bibitem[{Mennickent {et~al.}(1998)Mennickent, Patterson, O'Donoghue, Unda,
  Harvey, Vanmuster, \& Bolt}]{men98tucrt}
Mennickent, R.~E., Patterson, J., O'Donoghue, D., {et~al.} 1998, \apss, 262, 1

\bibitem[{Molnar \& Kobulnicky(1992)}]{mol92SHexcess}
Molnar, L.~A. \& Kobulnicky, H.~A. 1992, \apj, 392, 678

\bibitem[{Munari \& Zwitter(1998)}]{mun98CVspec5}
Munari, U. \& Zwitter, T. 1998, \aaps, 128, 277

\bibitem[{Noda {et~al.}(2002)Noda, Takeuti, Abe, Bond, Dodd, Hearnshaw, Honda,
  Honma, Jugaku, Kabe, Kan-ya, Kato, Kilmartin, Matsubara, Masuda, Muraki,
  Nakamura, Nankivell, Noguchi, Ohnishi, Reid, Rattenbury, Saito, Sato,
  Sekiguchi, Skuljan, Sullivan, Sumi, Watase, Wilkinson, Yamada, Yanagisawa,
  Yock, \& Yoshizawa}]{nod02MOALMC}
Noda, S., Takeuti, M., Abe, F., {et~al.} 2002, \mnras, 330, 137

\bibitem[{Nogami {et~al.}(1995)Nogami, Kato, Masuda, \& Hirata}]{nog95hvaur}
Nogami, D., Kato, T., Masuda, S., \& Hirata, R. 1995, \ibvs, 4163

\bibitem[{Nogami {et~al.}(1997)Nogami, Masuda, \& Kato}]{nog97sxlmi}
Nogami, D., Masuda, S., \& Kato, T. 1997, \pasp, 109, 1114

\bibitem[{O'Donoghue {et~al.}(1991)O'Donoghue, Chen, Marang, Mittaz, Winkler,
  \& Warner}]{odo91wzsge}
O'Donoghue, D., Chen, A., Marang, F., {et~al.} 1991, \mnras, 250, 363

\bibitem[{Osaki(1985)}]{osa85SHexcess}
Osaki, Y. 1985, \aap, 144, 369

\bibitem[{Osaki(1995)}]{osa95eruma}
---. 1995, \pasj, 47, L11

\bibitem[{Osaki(1996)}]{osa96review}
---. 1996, \pasp, 108, 39

\bibitem[{Osaki \& Meyer(2002)}]{osa02wzsgehump}
Osaki, Y. \& Meyer, F. 2002, \aap, 383, 574

\bibitem[{Patterson(1998)}]{pat98evolution}
Patterson, J. 1998, \pasp, 110, 1132

\bibitem[{Patterson {et~al.}(1998)Patterson, Kemp, Skillman, Harvey, Shafter,
  Vanmunster, Jensen, Fried, Kiyota, Thorstensen, \& Taylor}]{pat98egcnc}
Patterson, J., Kemp, J., Skillman, D.~R., {et~al.} 1998, \pasp, 110, 1290

\bibitem[{Patterson {et~al.}(1981)Patterson, McGraw, Coleman, \&
  Africano}]{pat81wzsge}
Patterson, J., McGraw, J.~T., Coleman, L., \& Africano, J.~L. 1981, \apj, 248,
  1067

\bibitem[{Podsiadlowski {et~al.}(2001)Podsiadlowski, Han, \&
  Rappaport}]{pod01amcvn}
Podsiadlowski, P., Han, Z., \& Rappaport, S. 2001, \mnras,
  \submitted\astroph{0109171}

\bibitem[{Richter \& Greiner(1995)}]{ric95ircom}
Richter, G.~A. \& Greiner, J. 1995, in Cataclysmic Variables, ed. A.~Bianchini,
  M.~Della~Valle, \& M.~Orio (\PublisherKluwer), 177

\bibitem[{Robertson \& Honeycutt(1996)}]{rob96htcas}
Robertson, J.~W. \& Honeycutt, R.~K. 1996, \aj, 112, 2248

\bibitem[{Shafter(1997)}]{sha97novarate}
Shafter, A.~W. 1997, \apj, 487, 226

\bibitem[{Starrfield(1999)}]{sta99novareview}
Starrfield, S. 1999, \PhR, 311, 371

\bibitem[{Starrfield \& Sparks(1987)}]{sta87novareview}
Starrfield, S. \& Sparks, W.~M. 1987, \apss, 131, 379

\bibitem[{Starrfield {et~al.}(2000)Starrfield, Truran, \&
  Sparks}]{sta00novareview}
Starrfield, S., Truran, J.~W., \& Sparks, W.~M. 2000, \NewAR, 44, 81

\bibitem[{Stellingwerf(1978)}]{PDM}
Stellingwerf, R.~F. 1978, \apj, 224, 953

\bibitem[{Uemura {et~al.}(2001)Uemura, Kato, Pavlenko, Baklanov, \&
  Pietz}]{uem01v725aql}
Uemura, M., Kato, T., Pavlenko, E., Baklanov, A., \& Pietz, J. 2001, \pasj, 53,
  539

\bibitem[{van~der Woerd {et~al.}(1988)van~der Woerd, van~der Klis, van
  Paradijs, Beuermann, \& Motch}]{vanderwoe88lateSH}
van~der Woerd, H., van~der Klis, M., van Paradijs, J., Beuermann, K., \& Motch,
  C. 1988, \apj, 330, 911

\bibitem[{Vogt(1980)}]{vog80suumastars}
Vogt, N. 1980, \aap, 88, 66

\bibitem[{Vogt(1983)}]{vog83lateSH}
---. 1983, \aap, 118, 95

\bibitem[{Warner(1985)}]{war85suuma}
Warner, B. 1985, in Interacting Binaries, ed. P.~P. Eggelton \& J.~E. Pringle
  (Dordrecht: D. Reidel Publishing Company), 367

\bibitem[{Warner(1995{\natexlab{a}})}]{war95book}
Warner, B. 1995{\natexlab{a}}, Cataclysmic Variable Stars (\PublisherCambridge)

\bibitem[{Warner(1995{\natexlab{b}})}]{war95suuma}
---. 1995{\natexlab{b}}, \apss, 226, 187

\bibitem[{Whitehurst(1988)}]{whi88tidal}
Whitehurst, R. 1988, \mnras, 232, 35

\bibitem[{Wood {et~al.}(1995)Wood, Naylor, Hassall, \&
  Ramseyer}]{woo95htcasXray}
Wood, J.~H., Naylor, T., Hassall, B. J.~M., \& Ramseyer, T.~F. 1995, \mnras,
  273, 772

\bibitem[{Woudt \& Warner(2001)}]{wou01v359cenxzeriyytel}
Woudt, P.~A. \& Warner, B. 2001, \mnras, 328, 159

\bibitem[{Yanagisawa {et~al.}(2000)Yanagisawa, Muraki, Matsubara, Abe, Masuda,
  Noda, Sumi, Kato, Fujimoto, Sato, Bond, Rattenbury, Yock, Kilmartin,
  Hearnshaw, Reid, Sullivan, Carter, Dodd, Nankivell, Rumsey, Honda, Sekiguchi,
  Yoshizawa, Nakamura, Sato, Kabe, Kobayashi, Watase, Jugaku, Saito, \&
  Koribalsky}]{yan00MOAcam2}
Yanagisawa, T., Muraki, Y., Matsubara, Y., {et~al.} 2000, \ExA, 10, 519

\bibitem[{Zhang {et~al.}(1986)Zhang, Robinson, \& Nather}]{zha86htcas}
Zhang, E.-H., Robinson, E.~L., \& Nather, R.~E. 1986, \apj, 305, 740

\end{thebibliography}
\end{document}